\documentclass[notitlepage,letterpaper,showpacs,preprintnumbers,amsmath,nofootinbib,amssymb,twocolumn, superscriptaddress, hyperref]{revtex4-1}

\pdfoutput=1
\usepackage{amssymb,amsmath,latexsym,mathrsfs}
\usepackage{graphicx} 
\usepackage{bm}
\usepackage[normalem]{ulem}
\usepackage{hyperref}
\usepackage{comment}
\usepackage{lipsum}
\hypersetup{
    colorlinks=true,       
    }   
\usepackage{xspace}
\usepackage{soul,xcolor}

\newcommand\be{\begin{equation}}
\newcommand\ee{\end{equation}}
\newcommand\bea{\begin{eqnarray}}
\newcommand\eea{\end{eqnarray}}

\newcommand\ie{{\it i.e.}~}
\newcommand\eg{{\it e.g.,}~}

\renewcommand\({\left(}
\renewcommand\){\right)}
\renewcommand\[{\left[}
\renewcommand\]{\right]}

\begin{document}

\setstcolor{red}

\title{The running of featureful primordial power spectra}

\author{Stefano Gariazzo}
\affiliation{Instituto de F\'isica Corpuscular (IFIC), CSIC-Universitat de Valencia,\\
Apartado de Correos 22085,  E-46071, Spain}

\author{Olga Mena} 
\affiliation{Instituto de F\'isica Corpuscular (IFIC), CSIC-Universitat de Valencia,\\
Apartado de Correos 22085,  E-46071, Spain}

\author{Victor Miralles}
\affiliation{Instituto de F\'isica Corpuscular (IFIC), CSIC-Universitat de Valencia,\\
Apartado de Correos 22085,  E-46071, Spain}

\author{H\'ector Ram\'irez}
\affiliation{Instituto de F\'isica Corpuscular (IFIC), CSIC-Universitat de Valencia,\\
Apartado de Correos 22085,  E-46071, Spain}

\author{Lotfi Boubekeur}
\affiliation{Universidad San Francisco de Quito USFQ, Colegio de Ciencias e Ingenier\'ias El Polit\'ecnico,\\
campus Cumbay\'a, calle Diego de Robles y V\'ia Interoce\'anica, Quito EC170157, Ecuador.}

\begin{abstract}
Current measurements of the temperature and polarization anisotropy
power spectra of the Cosmic Microwave Background (CMB) seem to indicate that the naive expectation for the
slow-roll hierarchy within the most simple inflationary paradigm may
not be respected in nature. We show that a primordial power spectra
with localized features could in principle give rise to the observed  
slow-roll anarchy when fitted to a featureless power spectrum. 
Future CMB missions have the key to disentangle  among
the two possible paradigms and firmly establish the slow-roll
mechanism as the responsible one for the inflationary period in the
early universe. From a model comparison perspective, and assuming that
nature has chosen a featureless primordial power spectrum, we find that,
while with mock Planck data there is only weak evidence against a
model with localized features, upcoming CMB measurements may provide
strong evidence against such a non-standard primordial power spectrum.


\end{abstract}
\twocolumngrid
\maketitle

\section{Introduction} 

Inflation is the most elegant and so far successful theory that is able to
provide the seeds for the structures we observe today in our universe
and solve the main problems of the standard Big Bang Cosmology
simultaneously~\cite{Guth:1980zm,Linde:1981mu,Albrecht:1982wi}. 
The most economical description of the inflationary paradigm is based
on the addition of a single new scalar degree of freedom, dubbed the inflaton, coupled to Einstein Gravity and slowly-rolling down a
potential. Models of inflation are usually tested by means of their predictions
for the standard inflationary observables, among which we have the
tensor-to-scalar ratio $r$, which characterizes the amplitude of the
primordial gravitational wave spectrum, and three parameters governing
the scale dependence of the power spectrum $\mathcal{P}_\zeta(k)$: the scalar
spectral index $n_s$, its running $dn_s/d\ln k$ and possibly 
the running of the running $n_{\rm{run,run}}$. For a recent appraisal of the 
constraining power of $dn_s/d\ln k$ and $n_{\rm{run,run}}$ in disentangling
different inflationary scenarios, see e.g.~Refs.~\cite{Escudero:2015wba,Munoz:2016owz}. 

The values of these parameters and their associated $68\%$
confidence level (CL) errors arising from the latest Planck 2015 temperature and polarization
TT,TE,EE+lowP~\cite{Ade:2015lrj} data are:
\begin{eqnarray}
n_s &=& 0.9586\pm 0.0056\,,\nonumber\\
dn_s/d\ln k &=& 0.009 \pm 0.010\,,\\
n_{\rm{run,run}} &=& 0.025\pm 0.013\,.\nonumber
\end{eqnarray}
There are very interesting and slightly suspicious aspects in the
measured values of the parameters governing the primordial power
spectrum. For instance, there is a mild preference for a positive
$n_{\rm{run,run}}\sim 10^{-2}$, while the standard single field slow-roll inflationary
paradigm typically predicts a {\it negative} one. But what is
more important and remarkable is the fact that, even if the current
errors on both $n_{\rm{run,run}}$ and $dn_s/d\ln k$ are still large to deduce any
strong conclusion, the mean values of these parameters do not seem to
follow the expected hierarchy within the simplest slow-roll
expansion. Namely, within this context, one would naively expect that $dn_s/d\ln k\simeq (n_s-1)^2$ and $ n_{\rm{run,run}}\simeq (n_s-1)^3$. These 
observational findings have previously motivated other works to
look for alternative inflationary models in which a different
hierarchy is expected, see Ref.~\cite{vandeBruck:2016rfv}.

Apart from the canonical single field slow-roll scenario, which will
lead to the standard power-law primordial power spectrum, there exist
a vast number of inflationary models in which the primordial power
spectrum possesses some features, see Ref.~\cite{Chluba:2015bqa} for
an extensive review. Examples of possible theoretical
scenarios in which a feature in $\mathcal{P}(k)$ may arise are, for instance,
models in which there are non-canonical kinetic terms in the
Lagrangian~\cite{Chen:2006nt}, where the value of the sound speed of 
the primordial curvature perturbation $c_s$ differs from the
value $c_s=1$ expected in the single-field slow-roll paradigm.
Other examples of featured models are those in which the sound speed varies 
with time~\cite{Chen:2005fe,Bean:2008na,Miranda:2012rm,Park:2012rh,Achucarro:2012fd,Gariazzo:2016blm}
or those governed by an inflaton potential with a sharp
step/feature~\cite{Adams:2001vc,Hunt:2004vt,Peiris:2003ff,Covi:2006ci,Ashoorioon:2006wc,Ashoorioon:2008qr,Jain:2008dw,Jain:2009pm,Mortonson:2009qv,
Hazra:2010ve,Benetti:2011rp,Adshead:2011bw,Adshead:2011jq,Bartolo:2013exa,Miranda:2013wxa,Miranda:2014wga,
Miranda:2015cea,Cadavid:2015iya,Benetti:2016tvm,Chen:2016vvw,Ballardini:2016hpi,GallegoCadavid:2016wcz},
or within the so-called axion monodromy scenarios~\cite{Silverstein:2008sg,McAllister:2008hb,Flauger:2009ab,Huang:2012mr,Easther:2013kla,
Meerburg:2014bpa,Meerburg:2014kna,Flauger:2014ana,Motohashi:2015hpa,Hazra:2016fkm} 

In this paper we focus on the possible interpretation of the current
cosmological data and of the forecasted constraints arising from future CMB missions in terms of a
featured primordial power spectrum shape.
Namely, the inflationary
mechanism realized in nature could be different from the usual
single-field slow-roll paradigm and the reconstructed values of the
$dn_s/d\ln k$ and $n_{\rm{run,run}}$ parameters could be hinting that.

The structure of the paper is as follows. In Sec.~\ref{sec:features} we present a simple, theoretically motivated, featured primordial power
spectrum.
In Section~\ref{sec:data}, we describe the method and the
cosmological probes exploited in our analysis. 
The present constraints on the usual slow-roll parameters, obtained when a
wrong assumption about the real shape of the power spectrum is made,
are described in Sec.~\ref{sec:constraints},
where we also explore the disentangling potential expected from future CMB missions.
We conclude in Sec.~\ref{sec:conclusions}.

\section{Features in the primordial potential: A toy model}
\label{sec:features}

The simplest realization of inflation arise from considering a sufficiently flat and smooth potential in which the slow-roll conditions are satisfied.
However, as previously stated, one could also consider models in which the inflationary potential exhibits features that modify the dynamics of the inflaton.
One particular option is to consider a class of well-motivated models in which the inflationary potential shows periodic modulations~\footnote{Features in the potential will usually break down the slow-roll approximation.
For sharp and high frequency features, developments have been made concerning the slow-roll techniques (see, \eg\cite{Motohashi:2015hpa,Miranda:2015cea} and references therein).}.
These ripples in the potential have the characteristic signature of enhancing the three-point correlation function of the primordial perturbations,
leading to a resonant primordial bispectrum and, thus, generating large non-Gaussianities~\cite{Chen:2008wn}. 

A subset of this class of resonant models arises naturally in the
framework of string theory. The so-called \emph{axion monodromy} model
makes use of the axion shift symmetry in order to address the problem of
Planck-suppressed terms in the effective Lagrangian, as well as
explaining the flatness of the inflaton potential. Furthermore, the
inflationary potential in these scenarios exhibits modulations
whose amplitude and frequency is given by the properties of the moduli
fields~\cite{McAllister:2008hb,Flauger:2009ab}. A simple realization
of a single-field monodromy model leads to an inflationary potential
of the form~\footnote{For more general realizations see e.g. Ref.~\cite{Flauger:2014ana}.}:
\begin{equation}
V(\phi)=V_0(\phi)+\Lambda^4\text{cos}\(\frac{\phi}{f}\)~.
\end{equation}
Here, $f$ is the axion decay constant and $\Lambda$ is the size of the modulations.

In general, a template for the primordial power spectrum within this class of resonant models reads as~\cite{Flauger:2014ana,Xu:2016kwz}
\begin{equation}
\mathcal{P}_\zeta(k)=\mathcal{P}_0(k)\left\{1+\frac{8f^{\text{res}}_{\text{NL}}}{\omega^2}~\text{cos}\[\omega~\text{ln}\(\frac{k}{k_*}\)\]\right\}~,
\label{eq:powerspectrum}
\end{equation}
where $f^{\text{res}}_{\text{NL}}$ is related to the amplitude of the
resonant non-Gaussianity, $k_*$ is the pivot scale (taken to be
constant and equal to $0.05$~$h$~Mpc$^{-1}$), $\omega$ is the
resonance \emph{frequency}~\footnote{The quantity \emph{frequency} $\omega$
  refers to the field inflation frequency divided by the Hubble
  parameter.}, which is related to the parameters of the
inflationary potential and $\mathcal{P}_0(k)$ is the primordial power spectrum, \ie, $\mathcal{P}_0(k)=A_{s_0}(k_*)
\(\frac{k}{k_*}\)^{n_{s_0}-1+\frac{1}{2!} \text{d}n_{s_0}/\text{d}\ln k\ln
  \(\frac{k}{k_*}\)+\frac{1}{3!} n_{\rm{run,run},0} (\ln
  \(\frac{k}{k_*}\))^2}$. 
Similar templates for the inflationary parameters can be derived from Eq.~(\ref{eq:powerspectrum}) as
\be
\begin{aligned}
n_s-1\equiv\frac{\text{d~ln}\mathcal{P}_\zeta(k)}{\text{d~ln}k}\simeq & ~(n_{s_0}-1)-\frac{8f^{\text{res}}_{\text{NL}}}{\omega}~\text{sin}\[\omega~\text{ln}\(\frac{k}{k_*}\)\]~,\\ \\
\frac{\text{d}n_s}{\text{d~ln}k}\simeq & ~\text{d}n_{s_0}/d\ln k-8f^{\text{res}}_{\text{NL}}~\text{cos}\[\omega~\text{ln}\(\frac{k}{k_*}\)\]~,\\ \\
n_{\rm{run,run}}\simeq &~n_{\rm{run,run},0} +8\omega f^{\text{res}}_{\text{NL}}~\text{sin}\[\omega~\text{ln}\(\frac{k}{k_*}\)\]~,\\
\end{aligned}
\ee
to be evaluated in $k=k_*$~\footnote{Notice that the perturbative
  terms in the expansion of Eq.~(\ref{eq:powerspectrum}) may become
  dominant if $\omega>1$ at a certain order.}.
The exercise that we perform along this study will consist on
interpreting Eq.~(\ref{eq:powerspectrum}) in terms of a featureless
primordial power spectrum:
\begin{widetext}
\be
\mathcal{P}(k)=A_s(k_*)
\(\frac{k}{k_*}\)^{n_{s}-1+\frac{1}{2!} \text{d}n_s/\text{d}\ln k\ln
  \(\frac{k}{k_*}\)+\frac{1}{3!} n_{\rm{run,run}} (\ln \(\frac{k}{k_*}\))^2}~.
\ee
\end{widetext}
after impossing that $\frac{\text{d}n_{s_0}}{\text{d~ln}k}$ and $n_{\rm{run,run},0}$ are
both zero. Therefore, the reconstructed values of $n_{s}$ and the
other remaining inflationary parameters could be, in principle, scale
dependent, as there is the
$\frac{8f^{\text{res}}_{\text{NL}}}{\omega^2}~\text{cos}\[\omega~\text{ln}\(\frac{k}{k_*}\)\]$
term in the primordial power spectrum which depends on $k$ and this
dependence should be mapped somewhere. However, in the absence of a
well-motivated and robust parameterization of such a dependence here we assume constant
values and explore the induced shifts in $n_{s}-1$,
$\text{d}n_{s}/d\ln k$ and $n_{\rm{run,run}}$. In other words, we
are interested in the shifts on the usual cosmological inflationary 
observables induced by our possible ignorance on the nature's
primordial power spectrum, resulting when fitting an axion monodromy
scenario to a featureless (but with non-zero $\text{d}n_{s}/d\ln k$ and
$n_{\rm{run,run}}$) power spectrum.

Analyzing the power spectrum, recent studies were able to
obtain the following constraint~\cite{Meerburg:2010ks}:
\begin{equation}
f^{\text{res}}_{\text{NL}}\lesssim10^{-3}{\omega^{5/2}}~.
\label{eq:constraint}
\end{equation}

Naively, for some allowed values of
the model parameters, one could expect to find
that there is an agreement between the predictions
of the primordial power spectrum in Eq.~(\ref{eq:powerspectrum}),
and the ones given by the standard slow-roll paradigm, taken into
account the non-zero values for the running of the spectral index, $dn_s/d\ln k$, and for the
running of the running, $n_{\rm{run,run}}$.
These two possibilities are
therefore expected to provide a similar fit to current data, as we
will illustrate in the following sections. Consequently, the
primordial power spectrum given by Eq.~(\ref{eq:powerspectrum})
should be regarded as a toy model.
Nevertheless
this simple model has been extensively proposed in the literature as a compelling
alternative to the slow-roll paradigm. We will use this simple model
as a working example throughout our study.

\section{Methodology and Cosmological data sets}
\label{sec:data} 
In order to quantify the viability of the resonant toy-model given by
Eq.~(\ref{eq:powerspectrum}), we  consider the Planck CMB satellite measurements of
the temperature and polarization anisotropies (the so-called TT, TE
and EE angular spectra), which extend up to a multipole
$\ell_{\mathrm{max}} = 2500$. We combine these measurements with
Planck low-multipole polarization data, ranging from multipoles
$\ell=2$ up to $\ell=29$.
We make use of the publicly available Planck likelihood
code~\cite{Aghanim:2015xee}, which also includes a number of nuisance
parameters, that we treat accordingly to Refs.~\cite{Ade:2015xua,Aghanim:2015xee}.
To derive the constraints on the different inflationary parameters, we make use of
the Boltzmann equations solver \texttt{CAMB} code \cite{Lewis:1999bs} and apply
Markov Chain Monte Carlo (MCMC) methods by means of the latest version
of the \texttt{CosmoMC} package~\cite{Lewis:2002ah}. As for current constraints, we consider an extended $\Lambda$CDM model
described by the following set of parameters: 
\begin{equation}\label{parameterPPS}
\{\omega_b,\omega_c, \Theta_s, \tau,\ln{(10^{10} A_{s})},n_s,r, dn_s/d\ln k, n_{\rm{run,run}}\}~,
\end{equation}
where $\omega_b\equiv\Omega_{b}h^2$ and $\omega_c\equiv\Omega_{c}h^2$
represent the physical baryon and cold dark matter energy densities, $\Theta_s$
is the angular scale of recombination,  $\tau$ is the reionization optical
depth,
$A_s$ is the normalization of the primordial power spectrum,
$n_s$ is the scalar spectral index, $dn_s/d\ln k$ and
$n_{\rm{run,run}}$ are the running and the running of the running of the spectral index.
The priors for these parameters are shown in Tab.~\ref{tab:priors}, both for the standard MCMC and the \texttt{PolyChord} analyses (see Sec.~\ref{sec:constraints}). 

\begin{table}
\begin{center}
\begin{tabular}{ccc}
\hline\hline
 Parameter & Prior &\texttt{PolyChord} prior\\
\hline
$\omega_b\equiv\Omega_{b}h^2$ &$0.005 \to 0.1$&$0.02\to 0.024$\\
$\omega_c\equiv\Omega_{c}h^2$ &$0.01 \to 0.99$&$0.1\to 0.14$\\
$\Theta_s$ & $0.5 \to 10$&$1.035\to 1.045$\\ 
$\tau$ &$0.01 \to 0.8$&$0.01\to 0.2$\\
$\ln{(10^{10} A_{s})}$& $2.7 \to 4$& $2.8 \to 3.4$\\
$n_{s}$  &$0.9 \to 1.1$&$0.9 \to 1.02$\\
$dn_s/d\ln k$  &    $-0.5 \to 0.5$&$-0.1\to0.1$\\
$n_{\rm{run,run}}$    &    $-0.5 \to 0.5$&$-0.1\to0.1$\\
\hline\hline
\end{tabular}
\caption{Uniform priors for the cosmological parameters considered in the present analysis.}
\label{tab:priors}
\end{center}
\end{table}

We shall also perform forecasted MCMC analyses to estimate the expected
constraining power of future CMB data in the context of featured
models, generating mock data for a cosmological model described by the
parameters above detailed, including $dn_s/d\ln k$ and $n_{\rm{run,run}}$. The
best-fit values for these parameters are chosen to be those detailed
in Ref.~\cite{Ade:2015lrj}. Then, we show the expectations from a Planck-like survey, to
compare the results with those obtained with real Planck data.
This could give us an appraisal of how much the
forecasted errors within an ideal scenario with perfect foreground subtraction
change when the true, real measurements are performed. For future CMB data, we
consider a COrE-like mission, following the specifications of
Ref.~\cite{Finelli:2016cyd}. 

Then, we repeat the same exercise above but assuming that nature has
chosen a featured primordial power spectrum. We
have assumed $\omega
=2.3501$ and $f^{\text{res}}_{\text{NL}}=0.0084$ in
Eq.~(\ref{eq:powerspectrum}) as benchmark values, as these values
provide a good fit to CMB measurements while still leading to a
featureful primordial power spectrum. These values are perfectly consistent with
the derived bounds on the amplitude of primordial power spectrum
perturbation in axion monodromy scenarios, see
e.g. Ref.~\cite{Meerburg:2014bpa}. Albeit there are other possible choices of $\omega$
and $f^{\text{res}}_{\text{NL}}$ satisfying Eq.~(\ref{eq:constraint}) which could also mimic the
observed values of the running and of the running of the running, we
restrict ourselves to illustrate one case, for the sake of
simplicity. The procedure is as follows. 
We first generate mock data assuming
a featured primordial power spectrum as the one given
by the resonant model within the axion monodromy scenario, see
Eq.~(\ref{eq:powerspectrum}).
Then, we fit this (mock data) model to a standard power spectrum
following the usual slow-roll expansion, to see whether a non-trivial 
primordial power spectrum with localized features could be mimicked by the
observed values of the running, $dn_s/d\ln k$, and of the
running-of-the-running, $n_{\rm{run,run}}$, of the scalar perturbations.
We present our main findings in the next section.

\section{Present and future constraints}
\label{sec:constraints} 
\begin{figure*}
\begin{tabular}{c}
\includegraphics[width=.87\textwidth]{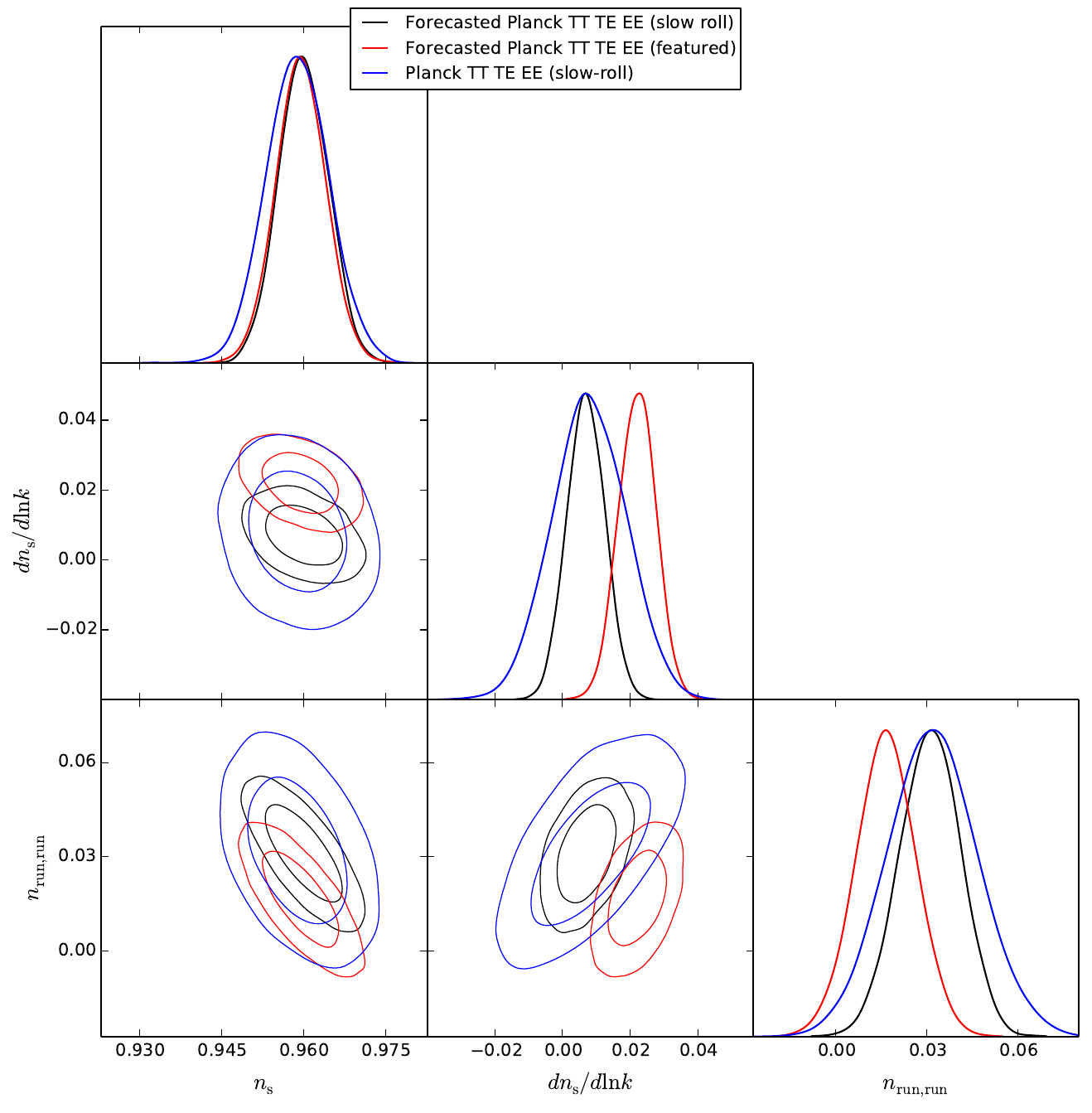}  \\
\end{tabular}
\caption{$68\%$ and $95\%$ CL in the two-dimensional 
($n_s$, $dn_s/d\ln k$), ($n_s$, $n_{\rm{run,run}}$) and ($dn_s/d\ln k$, $n_{\rm{run,run}}$)
planes as well as the one-dimensional posterior probability
distribution for the $n_s$, $dn_s/d\ln k$ and $n_{\rm{run,run}}$
parameters. The
black/blue curves illustrate the results resulting from an analysis
to Planck forecasted/real TT, TE and EE measurements, assuming the
standard slow-roll paradigm. The red curves denote the results when
the toy resonant model is (wrongly) fitted to Planck forecasted TT, TE and
EE measurements assuming slow-roll.}
\label{fig:fig1}
\end{figure*}

The present constraints are shown in Fig.~\ref{fig:fig1}, where we
show the $68\%$ and $95\%$ CL in the two-dimensional 
($n_s$, $dn_s/d\ln k$), ($n_s$, $n_{\rm{run,run}}$) and ($dn_s/d\ln k$, $n_{\rm{run,run}}$)
planes, as well as the one-dimensional posterior probability
distribution for each of the three parameters. 
We illustrate the allowed contours for three different analyses.
The
black (blue) curves illustrate the results from an analysis of Planck forecasted (current) TT, TE and EE measurements.
The red 
lines denote the results when the toy resonant model described 
in Sec.~\ref{sec:features}, describing axion monodromy inflation
scenarios, is fitted to  Planck forecasted TT, TE and
EE measurements assuming (incorrectly) the slow-roll paradigm.
It is very
important to notice that, albeit these results have been obtained from
a particular choice of the parameters governing Eq.~(\ref{eq:powerspectrum}) ($\omega
=2.3501$ and $f^{\text{res}}_{\text{NL}}=0.0084$)
to generate the mocks that afterwards are fitted to the slow-roll scheme, 
very similar results are obtained for a wide range of the toy-model
parameters, as previously explained.
This fact shows that, observationally, it is currently
very difficult to disentangle among featureless models and the plethora
of featured models described by the toy-model explored
here. Therefore, one can argue that the apparent slow-roll anarchy is due to
the fact that the primordial power spectrum is described by an axion
monodromy-like inflaton potential. This statement is further supported by the difference in the best-fit
$\chi^2$ values obtained for these two possibilities.
The modest 
value of $\Delta\chi^2\simeq 2$ obtained when fitting the resonant
toy model of Eq.~(\ref{eq:powerspectrum}) to the true underlying model, rather than to
the slow-roll scenario described by the $dn_s/d\ln k$ and $n_{\rm{run,run}}$,
suggests that the interpretation of current data in terms of a
primordial power spectrum with localized features is perfectly
plausible and compatible with the most recent CMB temperature and
polarization measurements. 
Additional constraints arising from bispectrum considerations do not change the
(current) findings above described, as the non-gaussianity parameter
$f_{\textrm{NL}}$ turns out to be negligibly small for the parameter
space of interest here. 

To further assess the fact that, with the present Planck data, an
underlying model with localized feautures in the primordial power
spectrum could be hidden in the form of a slow-roll anarchy
in which the slow-roll parameters do not respect the expected
hirarchical values, we have run the \texttt{CosmoMC} publicly
available code with the \texttt{PolyChord} nested sampler~\cite{Handley:2015fda}. This will
provide us the Bayesian evidence needed to compute the Bayes factor,
which will allow for a proper model comparison. Let us label by
$\mathcal{M}_0$ the model in which the (mock) Planck data is generated with a power
spectrum described by the slow-roll expansion and fitted to this very
same scenario. We instead refer to model $\mathcal{M}_1$ when the (mock) Planck data is generated with a featured
primordial power spectrum but it is fitted to a standard power-law model with
$dn_s/d\ln k$ and $n_{\rm{run,run}}$ different from zero. Then, the value we obtain
for the Bayes factor $|\ln B_{01}|=1.6$ indicates that there is only
weak evidence favoring $\mathcal{M}_0$ from Planck data (see
e.g.~Ref.~\cite{Trotta:2008qt}). This result further reinforces the findings quoted above. 

\begin{figure*}
\begin{tabular}{c}
\includegraphics[width=.86\textwidth]{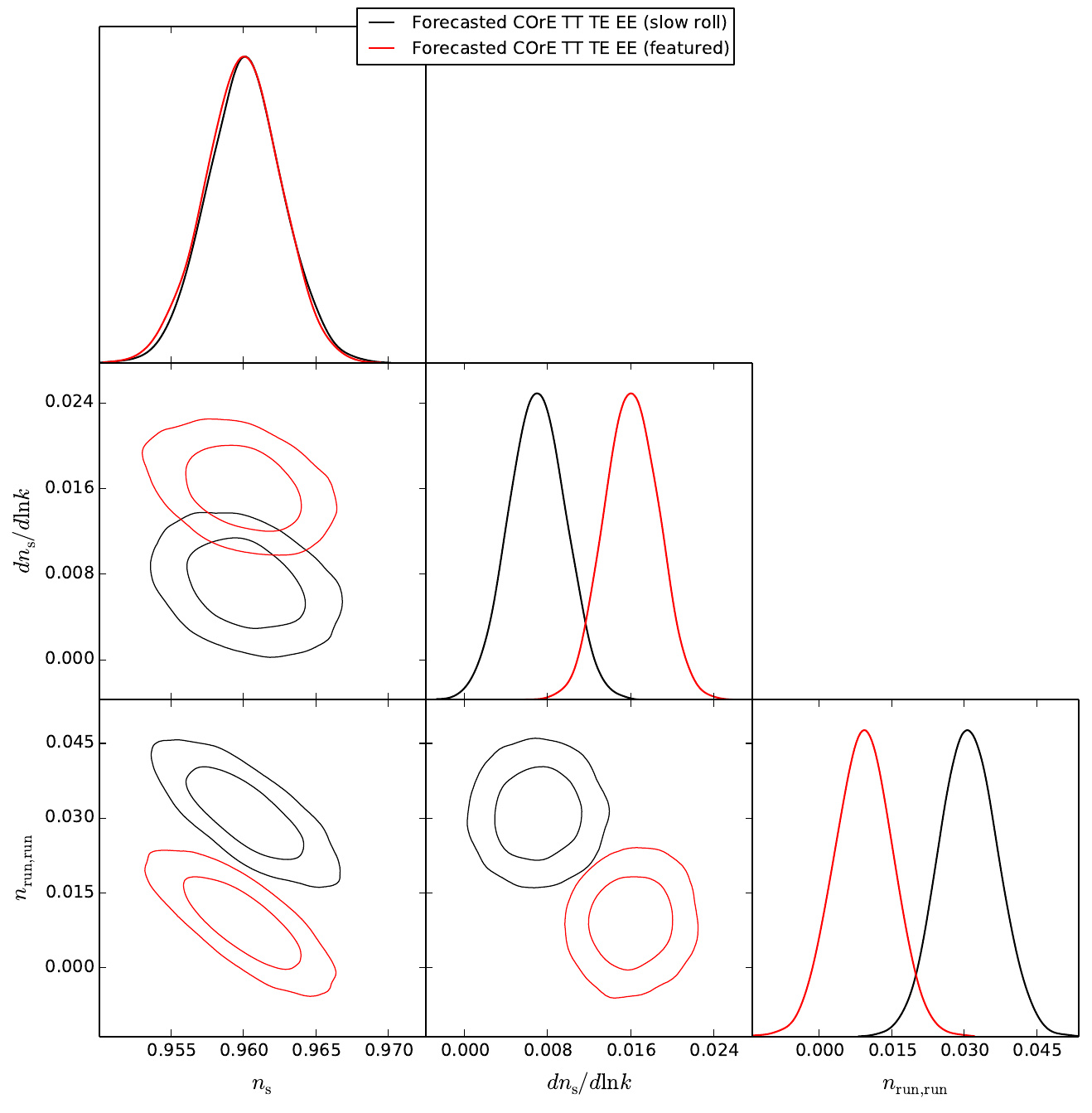} \\
\end{tabular}
\caption{As Fig.~\ref{fig:fig1} but for COrE mock data. Therefore,
  there are not equivalent curves to those shown in blue in Fig.~\ref{fig:fig1}.}
\label{fig:fig2}
\end{figure*}

Figure \ref{fig:fig2} shows the analogue of Fig.~\ref{fig:fig1} but
for mock CMB data generated accordingly to the future COrE mission
specifications~\cite{Finelli:2016cyd}. Even if in the COrE case the allowed contours
in the ($n_s$, $dn_s/d\ln k$), ($n_s$, $n_{\rm{run,run}}$) and ($dn_s/d\ln k$,
$n_{\rm{run,run}}$) planes are clearly separated, when compared to the
previous Planck case, it would be impossible to
single out the underlying nature's mechanism, as a priori one does
not know what the true model is. Notice also, from
Fig.~\ref{fig:fig1}, that the constraints from current Planck data are not as
 good as their forecasted values. However, one can not
 extrapolate this behaviour to the COrE case, as the impact from e.g. 
 systematics and foreground removals could look completely different
 in this case and the experience gained with Planck data
 cleaning will also help in matching the forecasted and real-data
 results. For the COrE case, we have also performed a proper model
 comparison analysis, as previously illustrated for the
 Planck case. The Bayes factor that we obtain in this
 case is $|\ln B_{01}|=7.2$, indicating that, if nature has chosen a
 featureless power spectrum, there should be strong
 evidence favoring this model. Nevertheless before truly assessing 
the nature's model for the generation of the primordial power
spectrum $\mathcal{P}_\zeta(k)$, independent measurements which firmly establish the existence 
of feautures in the primordial $\mathcal{P}_\zeta(k)$ are absolutely
required. In this regard, the $\mathcal{P}_\zeta(k)$ reconstruction from either galaxy
clustering data and/or from bispectrum measurements is a robust tool
which could provide a final answer.

\section{Conclusions}
\label{sec:conclusions}
Inflationary theories provide the most compelling solution to the
standard cosmological problems (horizon, flatness and generation of
primordial perturbations).
In its canonical version, inflation is
related to the existence of a scalar field, the inflaton, slowly
rolling down its potential.
This is known as the slow-roll  paradigm,
and leads to a hierarchy in the parameters governing the primordial
power spectrum's power-law.
Namely, the running $dn_s/d\ln k$ of the scalar
spectral index ($n_s$) and its running $n_{\rm{run,run}}$ are expected
to be second and third order in the slow-roll parameters, respectively. 
However, observationally, 
this hierarchy is not satisfied, with the current mean value of
$n_{\rm{run,run}}$ being larger than the corresponding one for
$dn_s/d\ln k$.
Even if errors are still very large to draw any
definite conclusion, one could look whether alternative inflationary
models predict a different hierarchy, closer to present
measurements~\cite{vandeBruck:2016rfv}.
In this regard, we have asked
ourselves whether this observed anarchy could be due to the fact
that the primordial power spectrum has some localized features, as in
theoretical scenarios with non-canonical kinetic terms, a time-varying
sound speed or within the so-called axion monodromy
scenarios.
We have focused here on this latter case, exploring a toy
model which reasonably describes the axion-monodromy inflationary predictions. 
Indeed, we have shown that when fitting mock Planck data
generated assuming a featured toy-model to a featureless power
spectrum, the values of the running of the scalar and of its running
can mimic the observed anarchy.
To reinforce our conclusions, we have also carried out a proper model
comparison analysis, and, assuming that nature has chosen a model with
the current mean values of $dn_s/d\ln k$ and $n_{\rm{run,run}}$, Planck
mock data show weak evidence when this model is compared to a model in
which a featured primordial power spectrum is fitted to the slow-roll
hierarchy one. A model comparison analysis in the COrE case will
provide strong evidence against the featured model, assuming that the
underlying true cosmology is a model with the standard power-law power
spectrum with values for the $dn_s/d\ln k$ and $n_{\rm{run,run}}$
parameters equal to their current best-fit values. Future CMB measurements, as those
expected to be carried out by the COrE satellite mission, have
the potential to disentangle among these two possibilities, 
even if the definite confirmation of the potential featureful nature of the
primordial power spectrum would also need the supporting proof from scale structure measurements and/or bispectrum data.

\begin{acknowledgments}
The authors thank E.~Giusarma and M.~Lattanzi for useful comments on the manuscript.
O.M. and H.R. are supported by PROMETEO II/2014/050, by the Spanish
Grant FPA2014--57816-P of the MINECO, by the MINECO Grant
SEV-2014-0398 and by the European Union's Horizon 2020
research and innovation programme under the Marie Sklodowska-Curie grant
agreements 690575 and 674896.
S.G. was supported by the Spanish grants FPA2014-58183-P, Multidark CSD2009-00064 and SEV-2014-0398
(MINECO), and PROMETEOII/2014/084 (Generalitat Valenciana).
\end{acknowledgments}
%

\end{document}